\documentstyle[aps,prd,preprint]{revtex}

\let\ssection=\section
\renewcommand{\section}{\setcounter{equation}{0}\ssection}

\begin{document}
\draft

\title{GENERALIZED GROSS--PERRY--SORKIN--LIKE SOLITONS}
\author{
Alfredo Mac\'{\i}as\footnote{E-mail: amac@xanum.uam.mx},\\ 
Departmento de F\'{\i}sica,\\
Universidad Aut\'onoma Metropolitana--Iztapalapa,\\
PO. Box 55-534, 09340 M\'exico D. F., MEXICO.\\
and\\
Tonatiuh Matos\footnote{E-mail: tmatos@fis.cinvestav.mx},\\ 
Departamento de F\'{\i}sica,\\
CINVESTAV--IPN,\\
PO. Box 14--740, 07000 M\'exico D. F., MEXICO}

\date{ \today}
%
\maketitle

\begin{abstract}
In this paper, we present a new solution for the effective theory of
Maxwell--Einstein--Dilaton, Low energy string and Kaluza--Klein theories,
which contains among other solutions the well known Kaluza--Klein
monopole solution of Gross--Perry--Sorkin as special case. We show also
the magnetic and electric dipole solutions contained in the general one.
\end{abstract}

%

\section{Introduction}
One of the exact solutions of the vacuum Einstein field equations in 
5-dimensional gravity is the Gross--Perry--Sorkin spacetime (GPS) \cite{GPS}, 
which is stationary, everywhere regular and without event horizon.
Actually, it represents a monopole, although there is no reason why the charge
carried out by the monopole should be labeled ``magnetic". It might equally 
well be deemed ``electric" and $A_{\mu}$ treated as a potential for the dual 
field $^* F_{\mu\nu}$. 

As it is well known \cite{GPS}, the monopoles, in addition to charge, are 
characterized by the topology of their spatial solutions. They carry one unit 
of Euler character, and therefore, one can construct stationary dipole 
solutions.
 
5D gravity is one example of the unified theories of electromagnetism and 
General Relativity. Einstein--Maxwell--Dilaton, Kaluza--Klein, and Low energy 
string theories are also examples of this kind of unified theories. 
Mathemetically their effective actions in four dimensions are very similar, 
they differ in that the value of the scalar dilatonic field coupling constant 
is in each case different. Thus we can write the four dimensional effective 
action for all of the above mentioned theories in the form:
\begin{equation} {\it S}=\int d^4x\sqrt{-g}[-R+2(\nabla\Phi)^2 
+ e^{-2\alpha\Phi}F_{\mu\nu} F^{\mu\nu}] \label{eq:ac} \end{equation}
\noindent
where $R$ is the Ricci scalar, $\Phi$ is the scalar dilaton field, 
$F_{\mu\nu}$ is the Faraday electromagnetic tensor, and $\alpha$ is the dilaton
coupling constant. For $\alpha=0$ we have the effective action of the 
Einstein-Maxwell-Dilaton theory, here, the scalar dilaton field appears 
minimaly coupled to electromagnetic one. $\alpha=1$ represents the Low energy 
string theory, where only the $U(1)$-vector gauge field has not been dropped 
out, and $\alpha = \sqrt{3}$ reduces the action (\ref{eq:ac}) to that of the 
5D Kaluza-Klein theory. As is well known, all of these theories 
unify gravity and electromagnetism. It is interesting to note that for the 
String and Kaluza-Klein theories, the electromagnetic field can not be 
decoupled from the scalar dilaton field.

The Gross--Perry--Sorkin solution has already been studied in \cite{GPS}, 
and recently  harmonic maps ansatz \cite{mat} has been applied to the action 
(\ref{eq:ac}) in order to find exact solutions of this kind to its 
corresponding field equations \cite{MNQ}. 

In this work we present two new classes of exact solutions of the 
(\ref{eq:ac}) associated 
field equations, for arbitrary values of the $\alpha$ coupling constant, the
solutions are written in terms of a harmonic map, in such a way that for 
spacial values of this harmonic map, the solution represents monopoles, 
dipoles, quadrupoles etc. If we choose the harmonic map in order to have 
monopoles, for the particular case $\alpha=\sqrt{3}$ it reduces just to the 
Gross--Perry--Sorkin solution. This new solitonic solution is the spacetime 
of a monopole for some values of the free parameter $\delta$ with $\alpha$ 
arbitrary. In general it represents a soliton spacetime
with a singularity at $r=2m$. Nevertheless, the influence of the scalar 
dilaton field is important only in regions near the sigunlarity.

The plan of the paper is as follows. In section 2 we review the harmonic map
ansatz. In Sec. 3 we review very briefly the Gross--Perry--Sorkin solution.
In Sec. 4 we present the new classes of solutions and its correspondig 
spacetimes 
in each of the mentioned theories. In Sec. 5 we discuss the results and 
present the conclusions.

\section{Harmonic Map Ansatz}
We begin by considering the Papapetrou metric in the following 
parametrization:
\begin{equation}dS^{2} = {1\over f}\lbrack e^{2k}(d\rho^{2} + d\zeta^{2}) 
+ \rho^{2} d\varphi^{2} \rbrack - fdt^{2} \label{PPP}
\end{equation}
The harmonic map ansatz supposes that all tems of the metric 
depend on a set of functions $\lambda^i,\ (i=1...p)$, such that these 
functions $\lambda^i$ fulfills the Laplace equation 
\begin{equation}
\Delta\lambda=(\rho \lambda_{,z})_{,\bar z} 
+ (\rho \lambda_{,\bar z})_{,z} = 0 \label{hmap} 
\end{equation}
where 
\begin{equation} 
z = \rho + i\zeta\label{eq:zet}.
\end{equation}
Thus the field equations derived from the Lagrangian (\ref{eq:ac}) 
reduce to equations in terms of the $\lambda^i$ functions. In general these 
equations are easyer to solve. The advantage of this method is that it is
possible to generate exact space times for each solution of the Laplace 
equation. 

Fortunately, the harmonic map determines the gravitational and the 
electromagnetic potentials in such a way, that we can choose them to have 
electromagnetic monopoles, dipoles, quadripoles, etc \cite{MNQ}.

In the Papapetrou parametrization, the field equations reduce to one equation
for $f$ 
\begin{equation}
\Delta lnf=e^{-2\alpha\Phi}{1\over\rho}f\ A_{\varphi,z}A_{\varphi,\bar z}
\label{feq}
\end{equation}
and to one for the function $k$
\begin{equation}
2k_{,z}=4\rho(\Phi_{,z})^2-e^{-2\alpha\Phi}{f\over\rho}(A_{\varphi,z})^2+  
\rho\ (\ln f_{,z})^2.
\label{keq}
\end{equation}
\noindent
with one equation for $k_{,\bar z}$ with $\bar z$ in place of $z$. Let us 
suppose that the components of the Papapetrou metric depend only on one 
harmonic map $\lambda$. Here we present two solutions of these field equations
(\ref{feq}) and (\ref{keq}). By solving the general field equations coming 
from the metric (\ref{PPP}) in terms of one harmonmic map $\lambda$
with no electromagnetic field at all, we arrive at a solution given by 
\cite{MNQ}
\begin{equation}
f = e^{\lambda}~~~;~~~~k_{,z} = {\rho \over 2}
(4\alpha^{2}a^{2} + 1)(\lambda_{,z})^{2} 
\end{equation}
with the following form for the scalar dilaton field:
\begin{equation}
 e^{2\alpha\Phi} = \kappa_{0}^2 e^{2\alpha^{2}a\lambda}~~~;
~~~a=const. 
\end{equation}
with $\lambda$ a harmonic map, $i.e.$ a solution of the equation (\ref{hmap}).

The field equation for the function $k$ is always integrable if 
$\lambda$ is a solution of the Laplace equation. (For more details of 
the method see \cite{mat} and \cite{MNQ}).

The second solution we want to deal with here, contains elecromagnetic field.
It is given by
\begin{equation}
f={1\over(1-\lambda)^{2\over 1+\alpha^2}},\ \ \ \ \  
k=0,\ \ \ \ \ A_{\varphi,z}=Q\rho \lambda_{,z},\ \ \  A_{\varphi,\bar z}=-Q
\rho \lambda_{,\bar z}
\label{sol2}
\end{equation}
and the corresponding form for the scalar dilaton field given by
\begin{equation}
e^{-2\alpha\Phi} = {e^{-2\alpha\Phi_0}\over(1-\lambda)^{2\alpha^2\over 1
+\alpha^2}}~~~.
\end{equation}

\noindent
where the magnetic charge is related with the scalar one by 
$Q^2={4e^{2\alpha\Phi0}\over 1+\alpha^2}$. This solution, contains among 
others, the GPS one as spacial case. 

In the next section we briefly review the Kaluza--Klein monopole.
 
\section{Gross--Perry--Sorkin Monopole}
The Kaluza--Klein monopole, known as Gross--Perry--Sorkin solution, represents
the simplest and basic soliton, it is a generalization of the self--dual 
euclidean Taub--Nut solution \cite{NTU}, and is described by the following
metric:
\begin{eqnarray}
ds^{2}=-dt^{2} + (1+{4m\over r})^{-1}(dx^{5} + 4m(1-cos\theta)d\varphi)^{2}
\nonumber\\ 
+ (1+{4m\over r})(dr^{2}+r^{2}d\theta^{2}+r^{2}sin^{2} \theta 
d\varphi^{2}) \label{eq:gps1} 
\end{eqnarray} 
where $(r,\theta,\phi)$ are polar coordinates. For $dt=0$ the Taub--Nut 
instanton is obtained. As it is well known, the coordinate singularity
at $r=0$ is absent if $x^{5}$ is periodic with period $16\pi m=2\pi R$ 
\cite{mis}, with $R$ the radius of the fifth dimension. 
Thus
\begin{equation}
m={\sqrt {\pi G}\over 2e}
\end{equation}
and the electromagnetic potential $A_{\mu}$ is that of a monopole:
\begin{equation}
A_{\varphi}=4m(1-cos\theta)
\end{equation}
and
\begin{equation}
{\bf B}= {4m{\bf r}\over r}
\end{equation}
The magnetic charge of the monopole is fixed by the radius of the 
Kaluza--Klein circle
\begin{equation}
g={4m\over \sqrt{16\pi G}}={R\over 2\sqrt{16\pi G}}={1\over 2e}\, .
\end{equation} 
Moreover, the mass of the soliton is given by
\begin{equation}
M={m\over G}
\end{equation} 
The Gorss--Perry--Sorkin soliton solutions are soliton solutions also of the 
effective four dimensional theory, for the four metric $g_{\mu\nu}$ and
a massless scalar dilaton field $\Phi$, as well, with
\begin{equation}
ds_{4}^{2}=-{dt^{2}\over \sqrt{1+{4m\over r}}} +  \sqrt{1+{4m\over r}}(
dr^{2}+r^{2}d\theta^{2}+r^2sin^2\varphi^2)
\end{equation} 
and
\begin{equation}
\Phi={\sqrt {3}\over 4}\ln{(1+{4m\over r})}
\end{equation}
Although this is a singular solution of the effective four dimensional theory
it is a perfectly sensible soliton. The singularity arises because the 
conformal factor $e^{2\alpha\Phi }$ is singular at $r=0$.

\section{Generalized Gross--Perry--Sorkin Solution}
In the Boyer-Lindquist coordinates:
\begin{equation}\rho=\sqrt{r^2-2mr}\sin\theta,\ \ \ \ \zeta=(r-m)\cos\theta
\, .\label{eq:blc} \end{equation} 
the group of metrics we want to deal with can be written as follows:
\begin{eqnarray}ds_{4}^{2} = (1 - \lambda)^{2\over(1 + \alpha^{2})}\lbrace 
\lbrack 1-{2m\over r}+{m^{2} sin^{2}\theta\over r^2}\rbrack 
[{dr^2\over 1-{2m\over r}}+r^2d\theta^2]\nonumber\\
+(1-{2m\over r})r^2sin^2\theta d\varphi^2\rbrace -{dt^2\over 
 (1 - \lambda)^{2\over(1 + \alpha^{2})}}\label{eq:gps}\end{eqnarray} 
\medskip
with the scalar dilaton field given by
\medskip
\begin{equation}e^{2\Phi} = {e^{2\Phi_{0}}\over (1 - \lambda)^{2\alpha\over
1 +\alpha^{2}}}\label{eq:dil}\end{equation}
\medskip

\subsection{Case $m=0$}

This corresponds to conformally spherically symmetric spacetimes.
Choosing $m=0$ in (\ref{eq:gps}), the metric reduces to:
\begin{equation}ds^{2} = (1 - \lambda)^{2\over 1 + \alpha^{2}}\lbrace 
dr^2+r^2d\theta^2+r^2sin^2\theta d\varphi^2\rbrace -{dt^2\over 
 (1 - \lambda)^{2\over 1 + \alpha^{2}}}\label{eq:sps}\end{equation} 
with $\lambda$ a solution of the Laplace equation:
\begin{equation} \Delta\lambda = (r^{2}\lambda_{,r})_{,r} + 
{1\over sin\theta}(sin\theta \lambda_{,\theta})_{,\theta}=0
\label{eq:lp} \end{equation}
and as is well known, the spherical armonics are all solutions of 
(\ref{eq:lp}). Moreover, it is easy to show, that it is possible to construct 
arbitrary monopoles, by performing the following identification:
\begin{equation} A_{3,z}=Q\rho\lambda_{,z}\end{equation}
\begin{equation} A_{3,\bar {z}}=-Q\rho\lambda_{\bar {z}}\end{equation}
with $z,~~\rho~~and ~~\zeta$ given by (\ref{eq:zet}) and (\ref{eq:blc})
respectively. The Gross--Perry--Sorkin is thus obtained by choosing
the electromagnetic potential $A_{3}=Q(1-cos\theta)$ and $\alpha=\sqrt{3}$, 
consequently 
\begin{equation}ds^{2}_{4}=(1+{4M\over r})^{1\over 2}
(dr^{2}+r^{2}d\theta^{2}+r^2sin^2\varphi^2)
-(1+{4M\over r})^{-{1\over 2}}dt^{2}\, ;\quad 
e^{2\Phi}={e^{2\Phi_{0}}\over (1+{4M\over r})^{3\over 2}} 
\, ,\label{eq:s4}\end{equation}
being $\alpha$ is the dilaton coupling constant. As mentioned in the 
introduction, for $\alpha = 0$ this is a solution in the framework of
the Einstein--Maxwell plus Dilaton theory, with the same $A_{3}$: 
\begin{equation}ds^{2}_{4}=(1+{4M\over r})^{2}
(dr^{2}+r^{2}d\theta^{2}+r^2sin^2\varphi^2)
-(1+{4M\over r})^{-2}dt^{2}\, ;\quad 
e^{2\Phi}=e^{2\Phi_{0}} 
\, ,\label{eq:em}\end{equation}
for $\alpha = 1$,~ it reduces to a low energy string theory solution, 
\begin{equation}ds^{2}_{4}=(1+{4M\over r})
(dr^{2}+r^{2}d\theta^{2}+r^2sin^2\varphi^2)
-(1+{4M\over r})^{-1}dt^{2}\, ;\quad 
e^{2\Phi}={e^{2\Phi_{0}}\over (1+{4M\over r})} 
\, .\label{eq:ls}\end{equation}
Finally, for $\alpha = \sqrt {3}$~~ a 5-D
Kaluza-Klein solution is obtained in the sense that in this particular case
\begin{equation}dS_{5}^{2} = {1\over I}dS_{4}^{2} + I^{2} (A_{\mu}dx^{\mu}+
dx^{5})^{2} 
\end{equation}
with
\begin{equation} I^{3}=e^{2\alpha\Phi} \end{equation}
This last fact suggests that in higher dimensional theories the 4-dimensional
part should be multiplied by a conformal factor related to the dilaton field
in order to be physically meaningful.

Further generalizations can be carried out by taking $\lambda = 
{qcos\theta\over r}$, $m=0$:
\begin{equation}ds^{2} = (1 - {qcos\theta\over r})^{2\over 1 
+ \alpha^{2}}\lbrace 
dr^2+r^2d\theta^2+r^2sin^2\theta d\varphi^2\rbrace -{dt^2\over 
 (1 - {qcos\theta\over r})^{2\over 1 + \alpha^{2}}}\, .\label{eq:sps1}
\end{equation} 
and the scalar dilaton field takes the form
\begin{equation}e^{-2\Phi} = {e^{-2\Phi_{0}}\over (1 
- {qcos\theta\over r})^{2\alpha\over 1 +\alpha^{2}}}\, ,\label{eq:dil4}
\end{equation}
with $\alpha= 0,1,\sqrt{3}$. This matric contains a magnetic dipole 
moment whose magnetical four potentential is 
$A_3=-{1\over 2\sqrt{2}}{sin^2\theta\over r}$.

\subsection{Case $m \neq 0$}

The same can be done with $m \neq 0$. The monopole solutions are obtained by 
choosing $\lambda = q ln(1-{2m\over r})$ (but now with gavitational mass 
${2mq\over 1+\alpha^{2}}$), with 
$A_{3}=-{1\over \sqrt{2}}Q(1-cos\theta)$, thus, the corresponding metric 
is given by 
\begin{eqnarray}ds_{4}^{2} = (1 - q ln(1-{2m\over r}))^{2\over 1 
+ \alpha^{2}}
\lbrace \lbrack 1-{2m\over r}+{m^{2} sin^{2}\theta\over r^2}\rbrack 
[{dr^2\over 1-{2m\over r}}+r^2d\theta^2]\nonumber\\
+(1-{2m\over r})r^2sin^2\theta d\varphi^2\rbrace -{dt^2\over 
 (1 - q ln(1-{2m\over r}))^{2\over 1 + \alpha^{2}}}\label{eq:gm}\end{eqnarray} 
\medskip
with the scalar dilaton field given by
\medskip
\begin{equation}e^{-2\Phi} = {e^{-2\Phi_{0}}\over (1 - q ln(1-{2m\over r}))^{2
\alpha\over 1 + \alpha^{2}}}\, .\label{eq:dil2}\end{equation}
\medskip
Moreover, in order to achive a magnetic dipole solution, we identify  
$\lambda = {qcos\theta\over(r - m)^{2} - m^{2}cos^{2}\theta}$, with the 
following electromagnetic potential: 
$A_{3}={1\over 2\sqrt{2}}q_{0} {(r-m)sin^{2}\theta\over (r-m)^{2}-
m^{2}cos^{2}\theta}$, which corresponds to a magnetic dipole field.
The metric is given by
\begin{eqnarray}ds_{4}^{2} = (1 - {qcos\theta\over (r - m)^{2} 
- m^{2}cos^{2}\theta})^{2\over 1 + \alpha^{2}}\lbrace 
\lbrack 1-{2m\over r}+{m^{2} sin^{2}\theta\over r^2}\rbrack 
[{dr^2\over 1-{2m\over r}}+r^2d\theta^2]\nonumber\\
+(1-{2m\over r})r^2sin^2\theta d\varphi^2\rbrace -{dt^2\over 
 (1 -{qcos\theta\over (r - m)^{2} - m^{2}cos^{2}\theta})^{2\over 1 
+ \alpha^{2}}}\label{eq:gps2}\end{eqnarray} 
\medskip
with the scalar dilaton field given by
\medskip
\begin{equation}e^{-2\Phi} = {e^{-2\Phi_{0}}\over (1 - {qcos\theta\over
(r - m)^{2} - m^{2}cos^{2}\theta})^{2\alpha\over 1 + \alpha^{2}}}
\label{eq:dil3}\end{equation}

This metrics are all asymptotically flat and also they are flat for $q=m=0$. 

\section{Outlook}

\section*{Acknowledgment}
This work was partially supported by CONACyT Grants No. 3544--E9311 and
No. 3672--E9312.

\end{document}